\documentclass{aa} 
\usepackage{graphicx}
\usepackage{natbib}
\usepackage{txfonts}
\begin{document}
   \title{Multiwavelength observations of a giant flare on
CN~Leonis\thanks{Based on observations
collected at the European Southern Observatory, Paranal, Chile, 077.D-0011(A) 
and on observations obtained with \emph{XMM-Newton}, an ESA science mission 
with instruments and contributions directly funded by ESA Member States and 
NASA.}\fnmsep
\thanks{Table 1 is only available in electronic form
    at the CDS via anonymous ftp to cdsarc.u-strasbg.fr (130.79.128.5) 
    or via http://cdsweb.u-strasbg.fr/cgi-bin/qcat?J/A+A/}}
   \subtitle{I. The chromosphere as seen in the optical spectra}

   \titlerunning{A giant flare on CN~Leo in multiwavelength}
   \authorrunning{B. Fuhrmeister et al.}

   \author{B. Fuhrmeister$^{1}$,
          C. Liefke$^{1}$, J. H. M. M. Schmitt$^{1}$, \and A. Reiners$^{2}$
          }

   \offprints{B. Fuhrmeister}

   \institute{$^{1}$Hamburger Sternwarte, University of Hamburg,
              Gojenbergsweg 112, D-21029 Hamburg\\
              $^{2}$University of G\"ottingen, Friedrich-Hundt-Platz 1,
              D-37077 G\"ottingen\\
              \email{bfuhrmeister@hs.uni-hamburg.de}
               }


 
\abstract
   {}
  {Flares on dM stars contain plasmas at very different temperatures and thus
   affect a wide wavelength range in the electromagnetic
   spectrum. While the coronal properties of flares are studied best in X-rays,
   the chromosphere of the star is observed best in the optical and ultraviolet ranges. Therefore,
   multiwavelength observations are essential to study flare properties
   throughout the atmosphere of a star.  }
   {We analysed simultaneous observations with UVES/VLT and \emph{XMM-Newton} of the
    active M5.5 dwarf CN~Leo (Gl 406) exhibiting a major flare. The optical
    data cover the wavelength range from 3000 to 10\,000 \AA.}
   {From our optical data, we find an enormous wealth of chromospheric 
    emission lines occurring throughout the spectrum. We identify a total of 1143 emission lines,
    out of which 154 are located in the red arm, increasing the number of observed
    emission lines in this red wavelength range by about a factor of 10. Here we
    present an emission line list and a spectral atlas.
    We also find line asymmetries for \ion{H}{i}, \ion{He}{i}, and \ion{Ca}{ii} lines.
    For the last, this is the first observation of asymmetries due to a stellar flare.
    During the flare onset, there is additional flux found in the blue
    wing, while in the decay phase, additional flux is found in the 
    red wing. We interpret both features as caused by mass motions. In addition to the lines,
    the flare manifests itself in the enhancement of the continuum throughout
    the whole spectrum, inverting the normal slope for the net flare spectrum. 
     }
   {}


   \maketitle
%

\section{Introduction}

Flaring is a commonly observed phenomenon on late-type stars. During a flare
event, large amounts of energy are released probably from magnetic field reconfigurations
and emitted over a wide range of the electromagnetic spectrum. Although white light flares are relatively rare for the Sun, they 
are quite common among M dwarf stars, because of their lower photospheric background emission in the optical.
Chromospheric emission lines react sensitively
to flares in amplitude, line width, line shape, and wavelength shifts. 
In addition, optical flares often have X-ray counterparts, although there is
no one-to-one relationship between X-ray and optical flares. During large flares, dM stars 
can show an increase in X-ray luminosity by factors up to 100 
\citep[e.\,g.][]{Guedel_ProxCen_1,Guedel_ProxCen_2}, and even greater magnitude increases 
have occasionally been reported in the optical \citep[e.\,g.][]{Eason, Hawley_Pettersen}.

The thermal properties of the coronal flare plasma can be diagnosed from the
emitted soft X-ray emission, while gyrosynchrotron radio and microwave emission typically
traces nonthermal particles in the corona. Additionally, emission lines in the ultraviolet mainly diagnose the transition region. Different wavelength ranges thus provide complementary information, 
so that multiwavelength observations of flares allow to study the contemporaneous reaction of different layers of the stellar atmosphere. 
A prototypical multiwavelength campaign was carried out
by \citet{deJager} for UV~Cet, which was simultaneously observed in X-ray, optical, and
radio. As a more recent example, \citet{UVCet_Xray_UV} observed a correlation between ultraviolet 
and X-ray flux for stellar flares.
Another recent multiwavelength campaign was aimed at the flare star EV Lac and
provided data in the radio, the optical, the ultraviolet, and in X-rays \citep{Osten_EV_Lac_flare, Osten_EV_Lac_quiescence}.
These data cover a huge radio flare and a number of optical and X-ray flares.
\citet{Osten_EV_Lac_flare} note that, for a flare in one wavelength region, counterparts in other wavelength regions are often missing. An example of a giant
flare on AD~Leo observed in the ultraviolet and optical is described by
\citet{Hawley_Pettersen}. AD~Leo was again observed in multiwavelength
by \citet{Hawley_Allred}, where a number of medium-sized flares are covered and
discussed. \citet{Smith} compare radio and X-ray flares on five active
M dwarfs and provide possible mechanisms for uncorrelated flares. Multiwavelength observations also
show that quiescent X-ray, optical, and radio emission strengths are no longer correlated for late M and early L dwarfs \citep{Berger1,Berger3,Berger2}.     
 
Our target star \object{CN Leo} (Gliese~406) is a well-known nearby flare star with a spectral type of M5.5
\citep{Reid} or M6.0 \citep{Kirkpatrick91}. \citet{Reiners} find a mean magnetic field of  $Bf \sim 2.2$ kG, exhibiting night-to-night and, with less certainty, intra-night
flux variations.  CN~Leo has previously been the target
of a multiwavelength campaign in December 2005, covering the optical (VLT/UVES) and X-ray regimes
(\emph{XMM-Newton}; \citet{fuhrmeister_liefke}). During these observations, CN~Leo once again
proved to be quite active, showing two medium-sized and a number of smaller
flares. Unfortunately, both larger flares occurred at the end of the optical
observations and were therefore not properly covered. Now, CN~Leo was again
observed in the framework of a similar multiwavelength campaign, during which a 
spectacular flare occurred. This paper
is the first in a series analysing the flare data, and it is structured as
follows: In Sect. 2 we describe  our observations obtained with UVES and
\emph{XMM-Newton}. In Sect. 3 we give an overview over the timing behaviour during
the flare. The wide variety of chromospheric emission lines is reported in
Sect. 4, while asymmetries found in the line shape of some emission lines are
described in Sect. 5. A discussion and our conclusions are presented in Sect. 6.


\section{Observations and data analysis}\label{data}

The multiwavelength observations reported in this paper were obtained
simultaneously with \emph{XMM-Newton} and ESO's Kueyen telescope equipped
with the Ultraviolet-Visual Echelle Spectrograph (UVES) on three half nights in 2006:
May, 19$^{th}$/20$^{th}$ (first night), 21$^{st}$/22$^{nd}$ (second night), and 
23$^{rd}$/24$^{th}$) (third night). We describe the optical data of 
the first night here
in detail, whereas the X-ray data will be described in paper II of this series \citep{paper2}.  

UVES is a cross-dispersed echelle spectrograph that was operated
using a dichroic beam splitter providing a blue arm spectrum (recorded
with a single CCD) and a red arm spectrum (recorded with two 
CCDs).\footnote{A detailed description of the UVES spectrograph is 
available under http://www.eso.org/instruments/uves/doc/} For the
blue arm, a standard setup with central wavelength at 3460~\AA\, was used,
while in the red arm a non-standard setup was used. The setup with
central wavelength 8600~\AA\, with blue end at 6600~\AA\, was slightly
blue-shifted to cover the H$\alpha$ line.
In our specific setup, used for the CN~Leo observations presented here, the 
spectral coverage was
between 3050~\AA\ to 3860~\AA\ in the blue arm and 6400~\AA\ to 10080~\AA\ in the red arm
with a small gap from 8190~\AA\ to 8400~\AA\ due to the CCD mosaic. 
We used a slit width of 1\arcsec\, resulting in a resolution of $\sim 40\,000$.
On May 19/20, the exposure times were 1000~s in the blue arm and 200~s
in the red arm. 
In the three half-nights, our observations resulted in 68, 24, and 89 spectra in the red 
arm, and 16, 4, and 11 spectra in the blue arm.   

The UVES spectra were reduced using the IDL-based {\tt REDUCE} reduction package
\citep{reduce}. The wavelength calibration was carried out using Thorium-Argon spectra
and resulted in an accuracy of $\sim 0.03$~\AA\, in the blue arm and 
$\sim 0.05$~\AA\, in the red arm. Absolute flux calibration was carried out 
using the UVES master response curves and extinction files provided by ESO. 
For fitting individual line profiles, we made use of the CORA program \citep{cora},
that was originally designed for flux measurements of high-resolution X-ray emission lines.

In addition, photometry with a time resolution of $\sim$~1~sec was obtained 
with the UVES exposuremeters, i.e., two blue and red photometers located in the two arms of 
the spectrograph. However, the exposuremeters are mainly used for calibration and engineering 
rather than for science, so they are not flux-calibrated. We can nevertheless evaluate their 
spectral efficiency, taking  into account the efficiency curves of all components in the optical 
path (i.\,e. dicroic and pre-slit filter) as provided by ESO\footnote{http://www.eso.org/instruments/uves/tools/}, 
and assuming the quantum efficiency of a typical GaAS semiconductor device for the photometer itself. 
For our instrumental setup, the red exposuremeter should thus show a spectral 
efficiency similar to the Johnson R filter, covering the wavelength range from 
$\sim 5900$--7100~\AA, with its maximum at 6150~\AA. The blue exposuremeter 
covers a broader wavelength range, from $\sim 3500$--5800~\AA, with a broad 
peak between 4600--5100~\AA, and thus comprises a mixture of typical B and V 
filters.

The \emph{XMM-Newton} satellite carries three co-aligned X-ray telescopes, each of which is 
equipped with one of the EPIC (European Photon Imaging Camera) MOS or PN 
instruments. 
These X-ray CCD detectors are photon-counting and thus provide
timing information at the subsecond level and medium-resolution spectral 
resolution. 
Two of the X-ray telescopes are additionally equipped with Reflection Grating 
Spectrometers (RGS) that provide high-resolution X-ray spectroscopy. The X-ray 
instruments are accompanied by the Optical Monitor OM, an optical/UV telescope 
that can be used with different filters in imaging or fast readout mode; the 
latter mode provides photometry with a time resolution of $\sim$~1~sec. All 
instruments are normally operated simultaneously.\footnote{More details on the 
instruments onboard \emph{XMM-Newton} can be found in the \emph{XMM-Newton} 
Users' Handbook, available at http://xmm.vilspa.esa.es/external/xmm$\_$user$\_$support/documentation/ uhb/index.html}

Some of the X-ray data has already been published \citep{coronal_explosion}, and we will give a 
detailed analysis of the X-ray data in a subsequent paper \citep{paper2}.
In this paper we focus on the Optical Monitor, which was operated in fast mode with the 
U-band filter covering wavelengths between 3100--3900~\AA, perfectly matching 
the spectral range of the blue arm UVES spectra. The observations are split 
into several exposures, each with an exposure time of 3200~s with gaps of 
about 300~s in between. We analysed the XMM-Newton data with the OM packages 
of the \emph{XMM-Newton} Science Analysis System (SAS) software, version 7.0. 

\section{Time resolved spectral response of the flare}\label{time}

\begin{figure}
\begin{center}
\resizebox{\hsize}{!}{\includegraphics{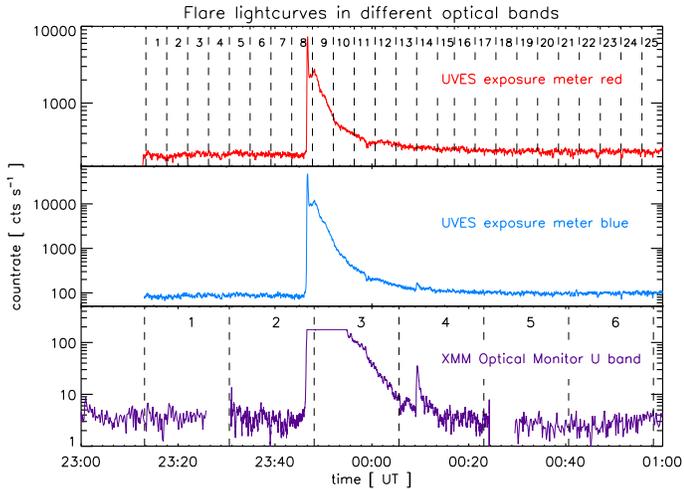}}
\caption{\label{lightcurve} Optical lightcurves of the large flare on CN~Leo. 
Top: UVES red arm exposuremeter. Middle: UVES blue arm exposuremeter. 
Bottom: \emph{XMM} Optical Monitor (cutted above count rates $>$200~cts s$^{-1}$). 
The vertical dashed lines in the red and U band plots mark the start times of 
the red and blue UVES spectra, respectively.}
\end{center}
\end{figure}

At the beginning of the first night of our observing campaign, a giant flare
occurred; the onset properties of this flare have been described in detail by \citet{coronal_explosion}.
The corresponding red, blue and the U-band lightcurves as derived from 
the two UVES exposuremeters and the OM, are 
shown in Fig. \ref{lightcurve}, where we also show the start times of the 
blue and red spectra. Note that the
spectra in the blue and red arms never start at the same time. Besides this 
large flare, several weaker flares occurred during the three half nights, 
which will be discussed in a separate publication. In this paper we concentrate on the emission observed from the giant flare.

The flare onset at UT 23:46:30 is seen as an enormous intensity increase in
all wavelength bands, i.\,e. in X-rays, in the U band, as measured with the OM, 
and with the blue and red UVES exposuremeters. 
During the first few seconds of the flare, a sharp peak in the optical and in soft
X-ray occurs. \citet{coronal_explosion} interpret this as the signature of a coronal explosion. After the first sharp peak, a broader
peak is visible in the optical UVES exposuremeter data, followed by an even
broader maximum in X-rays. The quasi-quiescent level after the flare
is 8 percent higher than the pre-flare level in the blue arm and 6 percent in the red arm. Also, the chromospheric emission lines remain enhanced after the flare.
A very similar time response 
of a large flare on UV~Cet with a sharp first peak, a broad second peak, and
the quiescent level after the flare higher than before was described by
\citet{deJager}.

At the first flare peak, the flux increased by a factor of about 550 in the blue UVES arm ,and by about 40 in the red arm compared to the pre-flare level. 
During the whole flare peak, the Optical Monitor entered saturation, i.\,e. the count rate in the U band exceeded the valid range for a reliable deadtime correction and flux conversion so that no reliable count rates can be derived. Additionally, the source extraction window suffers extremely from the so-called modulo-8 readout pattern, which disturbs the PSF of the OM. We thus do not consider the reconstructed shape of the lightcurve as trustworthy for count rates $>$200~cts~s$^{-1}$ and cut the lightcurve for
count rate values  exceeding this number.

The blue and red exposuremeter lightcurves are quite similar, and the 
decay after the broad second maximum is rather fast. In contrast, the decay 
time of the U-band lightcurve is much longer. Since the wavelength ranges 
covered by the exposuremeters contain far fewer chromospheric emission lines 
than the U-band, they mainly reflect the behaviour of the continuum, while the 
chromospheric emission lines dominate U-band lightcurve during the decay phase.

\subsection{Blue part of the spectrum}

\begin{figure*}[p]
\begin{center}
\resizebox{\hsize}{!}{\includegraphics{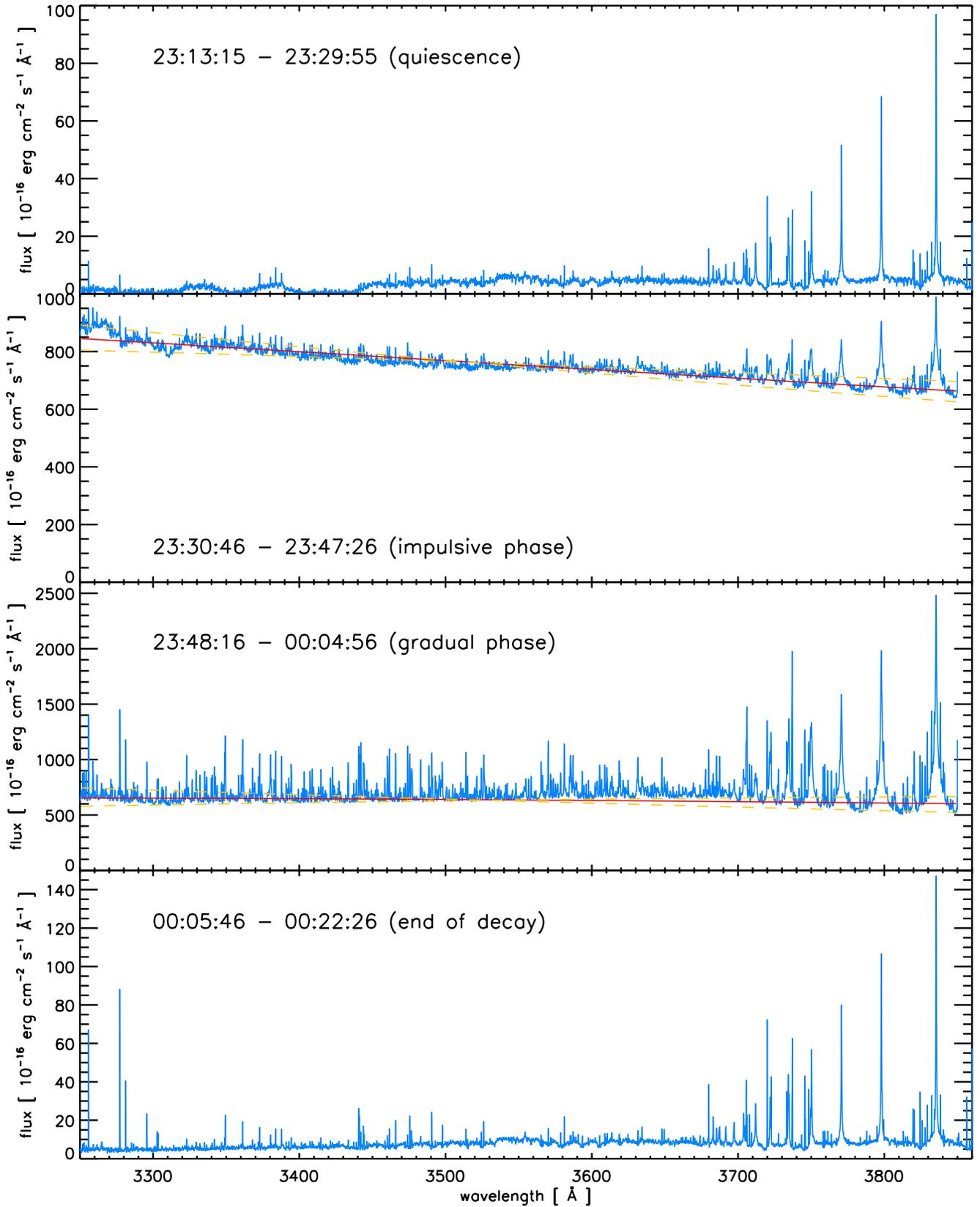}}
\caption{\label{blue} Sequence of blue UVES spectra no.~1 to 4 covering the flare (top to bottom). Note the different flux scales. Overlaid in red and orange
are different blackbody fits. }
\end{center}
\end{figure*}  

\begin{figure}
\begin{center}
\resizebox{\hsize}{!}{\includegraphics{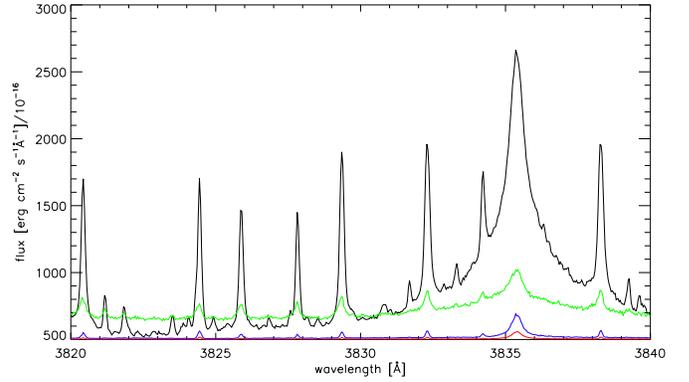}}
\caption{\label{h9} Spectral region around the Balmer line H9 in the blue arm. The evolution
of the hydrogen and of several metal lines can be seen. The blue/lowest
line denotes the pre-flare spectrum (no. 1). The green/light grey line
denotes the spectrum covering flare onset (no. 2). The black line and the red/grey
line denote the two following spectra (no. 3 and 4). The black spectrum (no. 3)
was used for the line identification.
}
\end{center}
\end{figure}

\begin{figure*}
\begin{center}
\includegraphics[width=18cm,height=24cm]{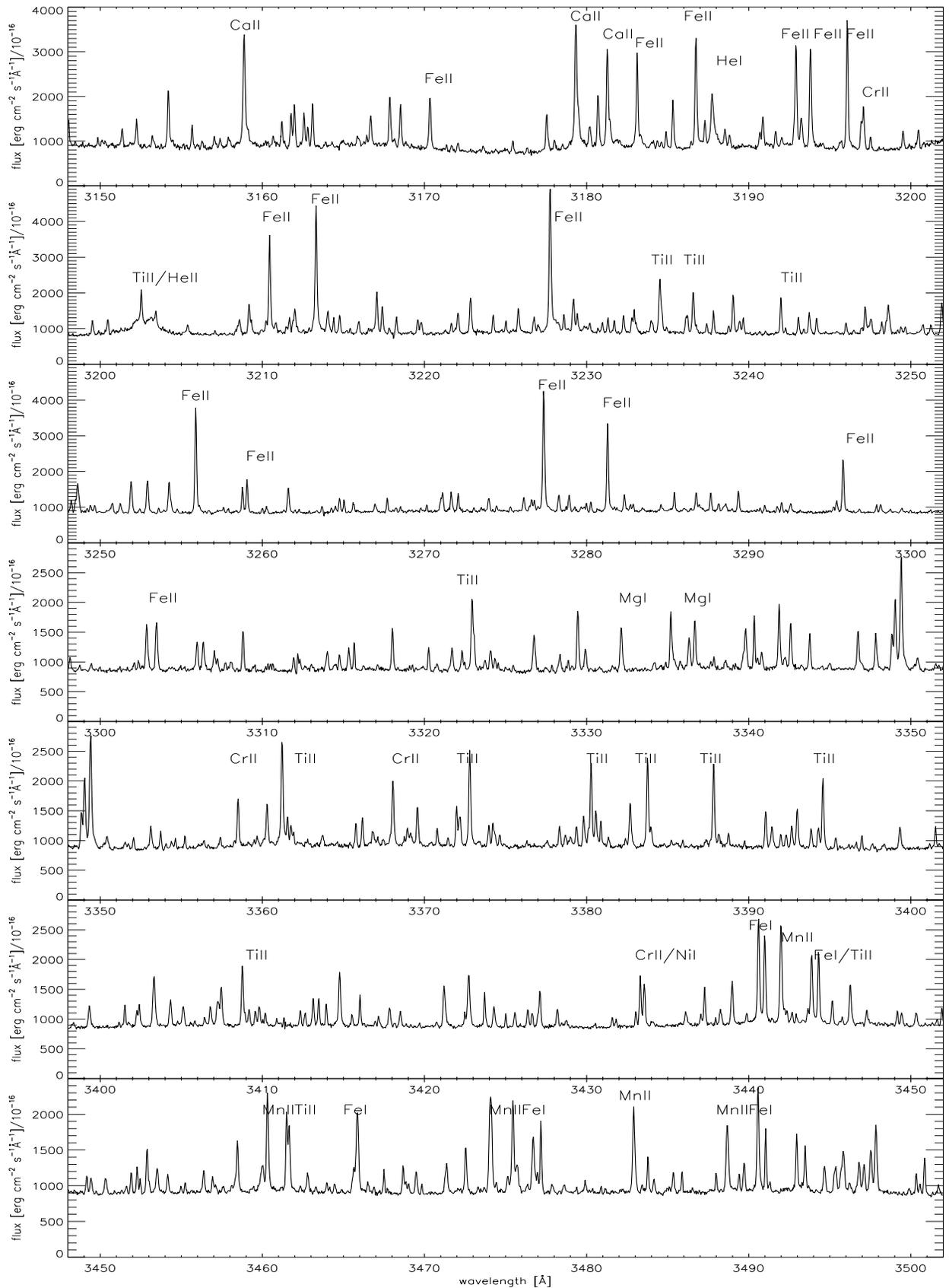}
\caption{\label{blueatlas} Spectral atlas of blue spectrum no. 3 from 3150 to 3500 \AA. 
Some important emission lines are identified, for other identifications see the emission line list.
}
\end{center}
\end{figure*}

\begin{figure*}
\begin{center}
\includegraphics[width=18cm,height=24cm]{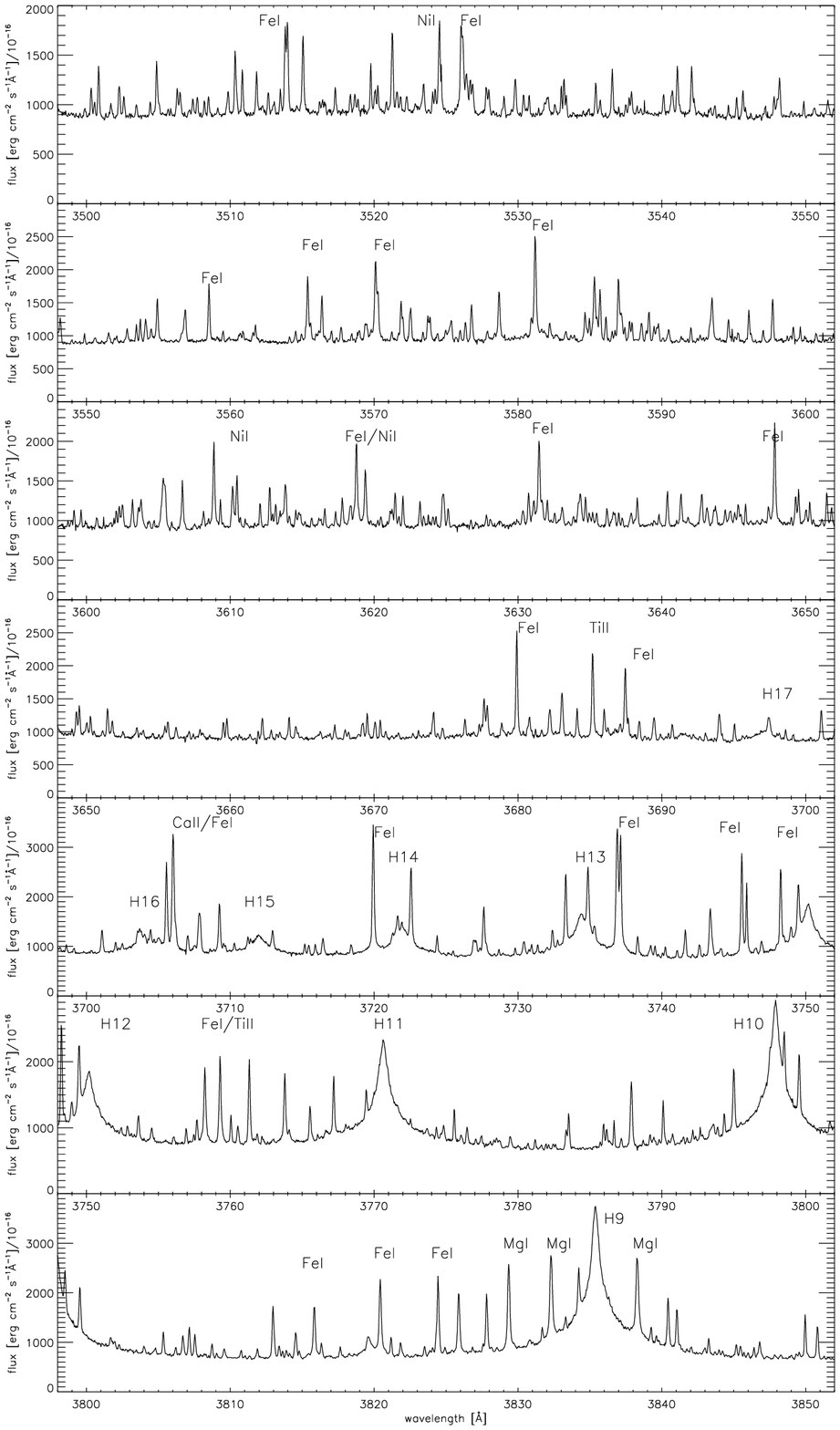}
\caption{\label{blueatlas2}Same as Fig. \ref{blueatlas} from 3500 to 3850 \AA.}
\end{center}
\end{figure*}

In the blue arm, the large flare is covered by spectra no. 2, 3, and 4. They are shown,
 together with the quiescent spectrum no.~1, in Fig~\ref{blue}. The last 
$\approx 60$ seconds of exposure no.~2, taken from UT 23:30:46 to 23:47:26, cover the 
impulsive phase of the flare; the contribution of quiescent emission in 
exposure no.~2 should be negligible. The spectrum is dominated by blackbody
emission (see section \ref{blackbody}). The Balmer 
lines have gained in amplitude and developed very broad wings. Many other 
chromospheric emission lines start to show up as well, but they are totally outshone by the blackbody
radiation during the first
$\approx 60$ seconds of the flare. Therefore, the true wealth of 
emission lines becomes visible only in spectrum no. 3, covering  
the broader peak of the flare from 23:48:16 to 00:04:56. In spectrum no. 4,
most of the emission lines show still increased emission, while in spectrum
no. 5, a new quiescent level is reached. Simultaneously, the flux reaches the quiescent 
level around 
UT 00:30 as shown in the lightcurve in Fig.~\ref{lightcurve}.  
 A detailed example of the evolution
of a typical part of the blue spectrum covering the H$_{9}$ line is shown in
Fig.~\ref{h9}. An atlas of spectrum no. 3, which was used to create an emission line list,
can be found in Figs. \ref{blueatlas} and \ref{blueatlas2} published 
as online material. 

\subsection{Red part of the spectrum}

In the red arm, the very first seconds of the flare are covered by spectrum no.~8, 
taken from
UT 23:43:20 to UT 23:46:40. In this spectrum, a slightly higher level of continuum 
emission is already visible, especially in the bluer part of the spectrum around 
H$\alpha$.  
Some lines, e.\,g. the \ion{He}{i} line at 6678 \AA, are starting to go
into emission, while others are not. This should partly be due to a lack of
sensitivity since only a rather
short time interval of the flare is covered, which prevents weak lines from
showing up against the continuum. However, this may also be due to non-equilibrium
conditions during the first few seconds of the flare, since lines formed at different
atmospheric depths should react on different timescales. 

\begin{figure*}[p]
\begin{center}
\includegraphics[width=18cm,height=24cm,bbllx=60,bblly=45,bburx=470,bbury=710]{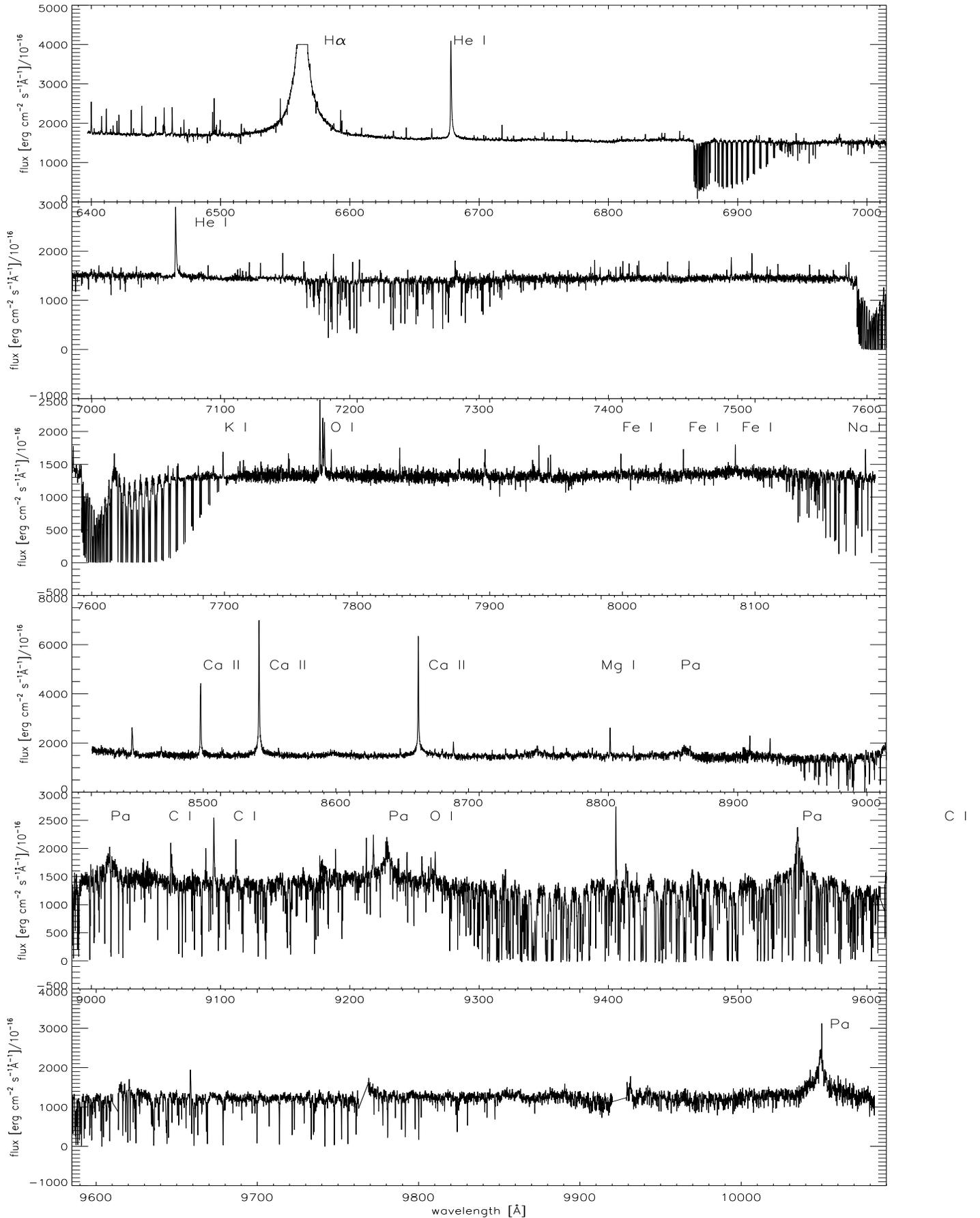}
\caption{\label{redall} Difference (flare-only) spectrum 
of the red arm mosaic of spectrum no.~9 and 1 (same as in Fig. \ref{redseq}). 
A wide variety of emission lines can be noted, which is
totally atypical for this wavelength range. Also the slope is inverted 
compared to the quiescent spectrum.
}
\end{center}
\end{figure*}

\begin{figure*}[!h]
\begin{center}
\resizebox{\hsize}{!}{\includegraphics{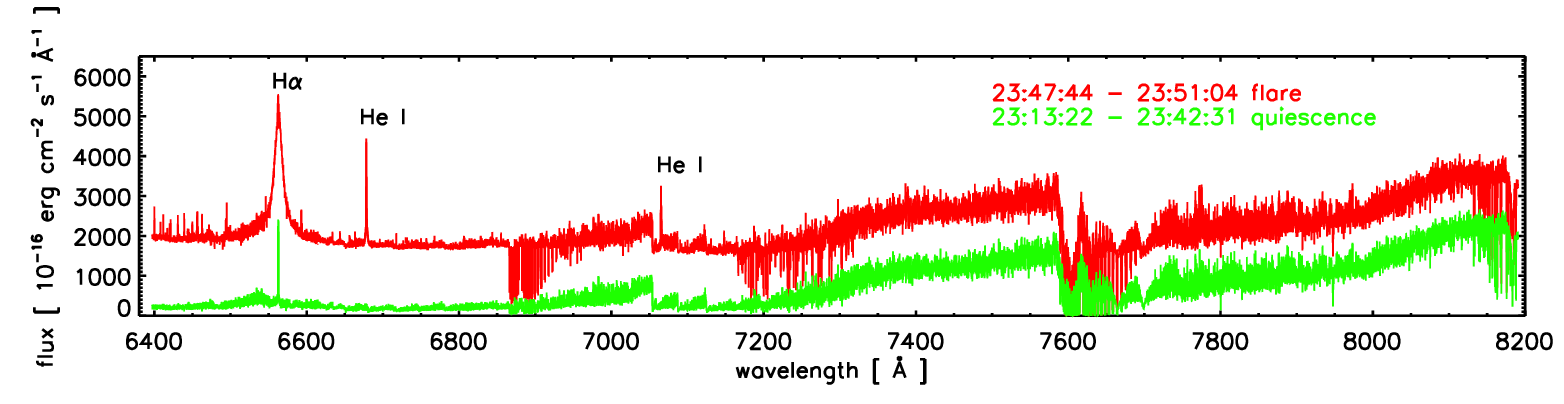}}
\resizebox{\hsize}{!}{\includegraphics{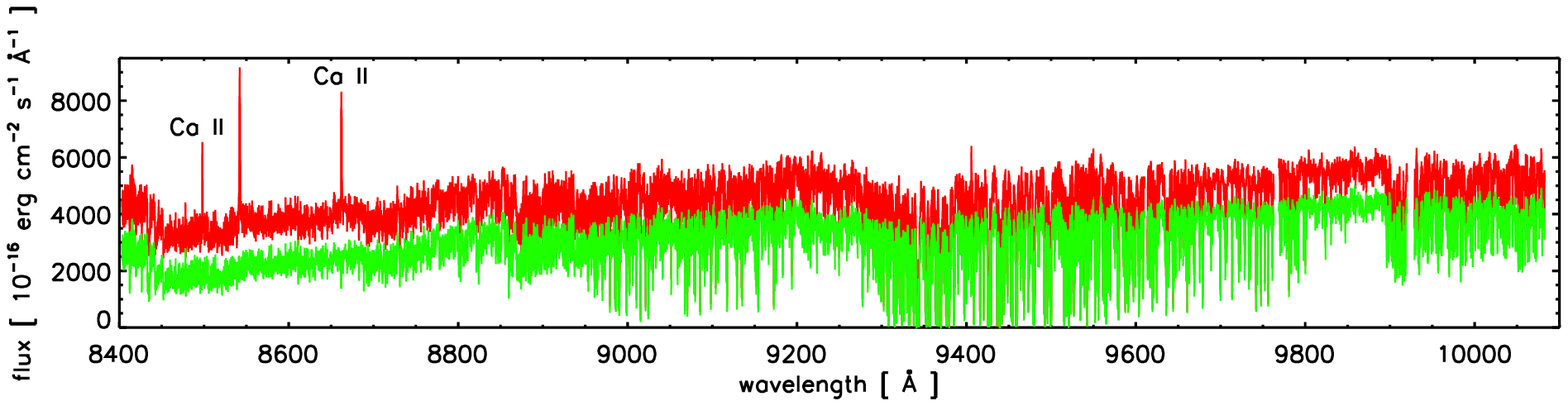}}
\caption{\label{redseq} UVES red spectra in quiescence (spectrum no.~1 in green/grey)
and in flare (spectrum no.~9 in red/black).}
\end{center}
\end{figure*}

The consecutive spectrum no.~9 starts at
UT 23:47:37. Here, the continuum enhancement reaches a maximum and
is clearly seen throughout the whole red spectrum. The net flare spectrum, i.\,e.
with the quiescent flux subtracted, clearly shows an inverse slope with the
flux increasing towards the blue end. 
This effect has already been described for a similar wavelength range for flares on 
2MASS~J0149090+295613 \citep[M9.5,][]{2Massflare} and 2MASS~J1028404-143843 
\citep[M7,][]{Schmidt}. Moreover, spectrum no.~9 shows a wide variety of emission lines never observed before in this spectral range. We show the flare spectrum no.~9 with spectrum no.~1 subtracted in Fig.~\ref{redall}, and the two spectra themselves in Fig.~\ref{redseq}. 

\begin{figure}[!h]
\begin{center}
\includegraphics[width=8cm]{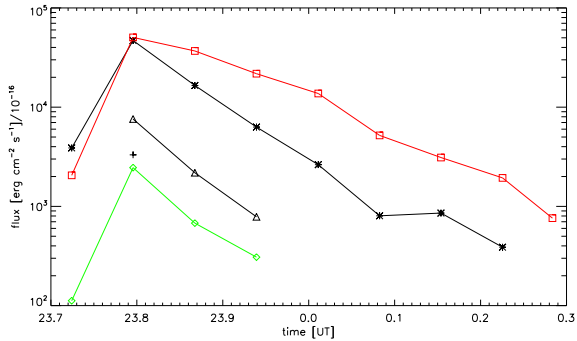}
\caption{\label{line_flux} Timing behaviour of line fluxes with quiescent flux removed 
for some important
emission lines. Red squares denote the \ion{Ca}{ii} line at 8542 \AA\, (narrow and
broad component; line core is saturated during flare peak),black asterisks denote 
the \ion{He}{i} line at 6678 \AA\, (narrow and broad component), black triangles
denote the \ion{O}{i} line at 7774 \AA, green diamonds denote the emission core of the
\ion{K}{i} line at 7699 \AA, and the black cross the line flux of the \ion{Na}{i} line, which
is elevated only during flare peak. 
}
\end{center}
\end{figure}

The emission
lines decay on very different timescales. For example, emission of the
\ion{Ca}{ii} triplet
lines decays very slowly and can be noted until spectrum no. 16, while the
emission core of \ion{K}{i} is undetectable after spectrum no. 11. The same applies
to the net flux of the three \ion{O}{i} lines around 7772 \AA, while the three \ion{O}{i}
lines around 9262 \AA\, are only found in the flare spectrum no. 9. 
The timing behaviour of the line flux is shown in Fig. \ref{line_flux} for some important
emission lines.

The continuum emission stays enhanced until about UT 00:15 (spectrum no. 15) in the blue part of the
red arm, while the red part of the red arm is only elevated in spectra 
no. 8, 9, and 10. 

\subsection{Temperature evolution}\label{blackbody}

The well-defined slope of the flare continuum in our spectra suggests a blackbody origin.
 We fitted temperature and emitting 
area to the smoothed blue spectra no.~2 and 3 after removing the emission 
lines. In both cases, the contribution of the quiescent atmosphere was 
neglected. The fit gives temperatures of $\sim 11\,300$~K for spectrum no.~2 
\citep[also discussed in][]{coronal_explosion} and of $\sim 9100$~K
for spectrum no. 3; however these temperatures correspond to 
averaged values during the exposure (i.\,e. for the flare onset during the 
first minute of the flare in spectrum no.~2, and over the whole exposure time 
of 1000~s during the broader peak in spectrum no.~3), since the thermal 
continuum emission can be expected to undergo a rapid decrease in temperature 
and flux. Figure~\ref{blue} shows the two best-fit blackbodies as straight (red) 
lines; blackbody fits with fixed temperatures of 10\,000~K and 15\,000~K in 
spectrum no.~2 and with 7000~K and 10\,000~K in spectrum no.~3 are also shown 
for comparison as dashed (orange) lines. Thus the uncertainties of the 
temperature fit can be estimated not to exceed 1000~K. The resulting sizes of 
the emitting area are then 1--$10 \cdot 10^{18}$\,cm$^2$. 
In spectrum no.~4, the continuum level has nearly decreased to
quiescent levels.

The same procedure can be applied to the red arm spectra no.~8, 9, and 
10, with the quiescent flux subtracted. Since for 
temperatures around 10\,000~K the wavelength range covered by the red arm 
spectra is located far in the Rayleigh-Jeans part of the blackbody spectrum, 
the derived temperatures have rather large errors, which we estimate to reach up to
5000~K. Additionally, the true continuum level at the 
position of the strong telluric absorption bands is difficult to assess. 
It is thus  appropriate 
to restrict the analysis to the short-wavelength part of the spectrum.
The red flare spectrum no.~8, covering the impulsive phase of the flare, 
can be fitted with a temperature of $\sim 27\,700$~K (and with $\sim 19\,600~$K
considering both parts of the red arm).
For the red spectrum no.~9 we determined a temperature of $\sim 5600$~K.
 Spectrum no.~10 gives  temperatures 
of 2600 to 3200~K, i.\,e. the flare plasma has cooled to  CN~Leo's quiescent 
photospheric temperature of $\approx 2900$~K \citep{Pavlenko, Fuhrmeister}. 
The values derived for the emitting area are consistent with those from the 
blue spectra alone. 

Putting the derived temperatures in the right chronological order, one starts
with red spectrum no.~8 (covering the first few second of the flare) 
with probably more than 20\,000~K, then followed by the blue spectrum
no.~2 (covering the first minute) with about 11\,000~K. In the subsequent
blue spectrum no.~3 and red spectrum no.~9, the measured temperature is rapidly
dropping from 9100~K to 5600~K, reaching the quiescent 
photospheric temperature for red spectrum no.~10.  

The optical flare
continuum emission can most likely be attributed 
to photospheric plasma that is reprocessing UV and EUV line emission, which itself was 
induced by heating from nonthermal particles \citep[as discussed by][where they confirm the 
blackbody hypothesis from photometric UV observations of flares on AD~Leo]{Hawley_Fisher}. Several 
spectra covering flares on late M dwarfs can be found in the literature; however, 
they typically cover only the spectral range of H$\alpha$ and longwards in wavelength, 
introducing large uncertainties into a possible blackbody fit. Our blue spectra allow us to 
assess the blackbody temperature from the slope of the spectra with reasonably small 
uncertainties. Together with the different temporal coverage of the flare in the red 
arm, we can conclude that the plasma is initially heated to temperatures probably 
greater than 20\,000~K and cooling rapidly afterwards, reaching already the photospheric level 
when the chromospheric line emission is still seen near its maximum.

\section{Identification of chromospheric emission lines}\label{lines}

For the flare spectra, we point out the wealth of chromospheric emission lines.
Because many of the lines, especially in the red part of the
spectrum, were not observed in emission before, we provide
an emission line list for the flare spectrum. 

\subsection{General information on the line catalogue}

\begin{figure}
\begin{center}
\includegraphics[width=8cm]{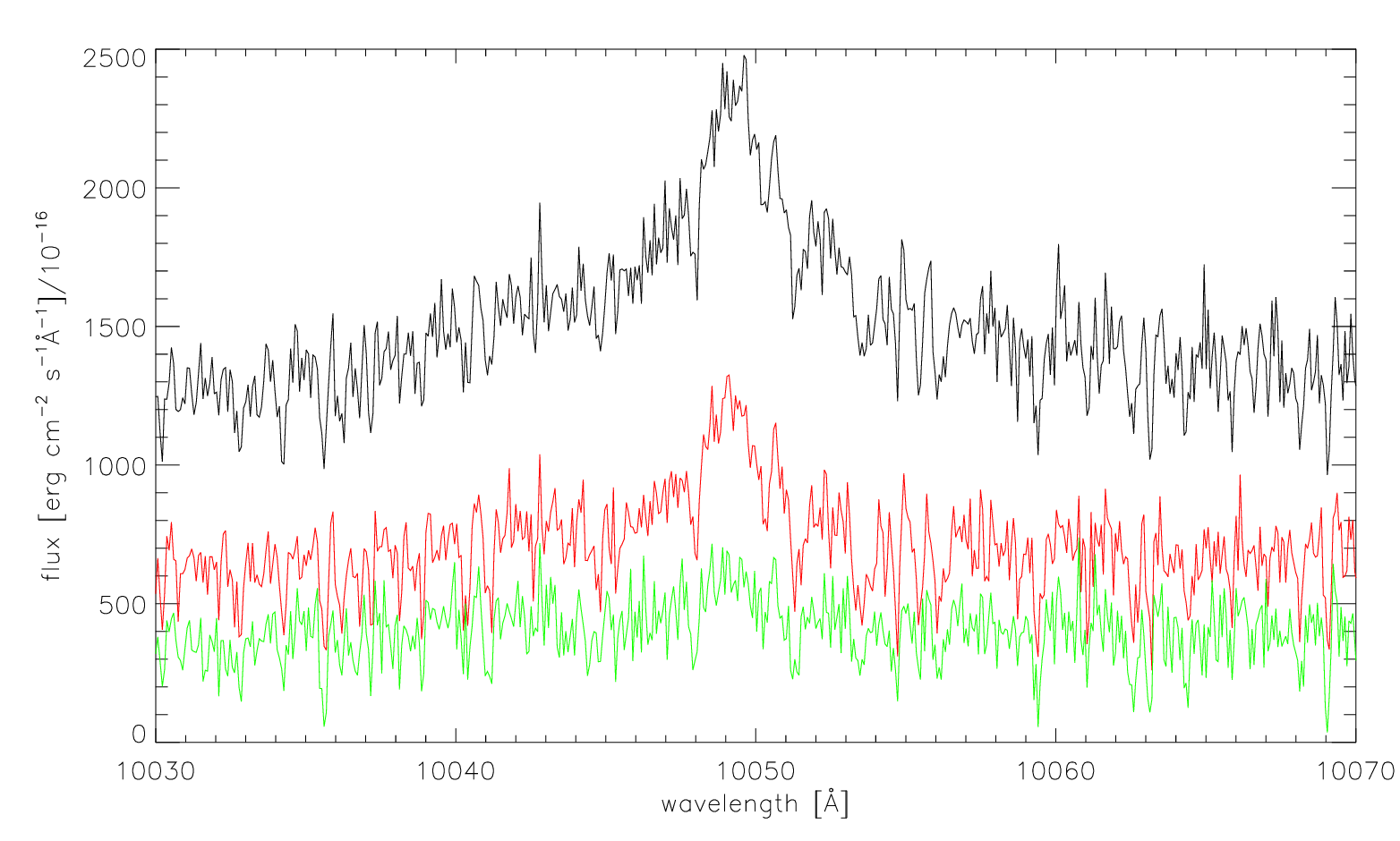}
\caption{\label{paschen} Paschen (P7) line at 10049 \AA\, with a cosmic removed
at 10049.972 \AA. The black line corresponds to the red flare-only
spectrum no. 9. Also shown are the two consecutive flare-only spectra, where the
line quickly decays.
}
\end{center}
\end{figure}

\begin{figure}
\begin{center}
\includegraphics[width=8cm]{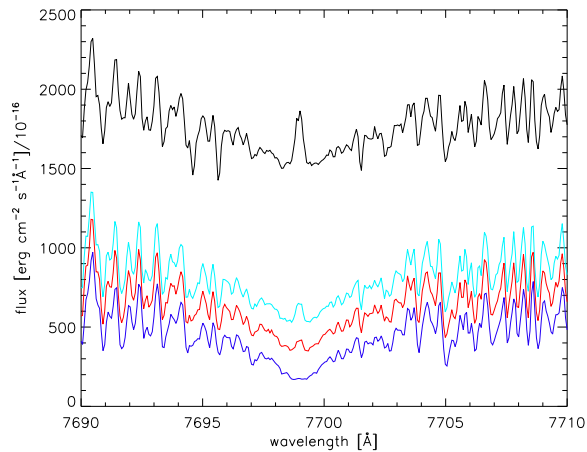}
\caption{\label{kline} \ion{K}{i} line at 7699 \AA. The quiescent spectrum
no. 1 is shown in blue (lowest line). The black line denotes the flare spectrum no. 9,
the turquoise/light grey and red/grey lines denote the two following spectra.
Note the emission core during the flare. 
}
\end{center}
\end{figure}

\begin{figure}
\begin{center}
\includegraphics[width=8cm]{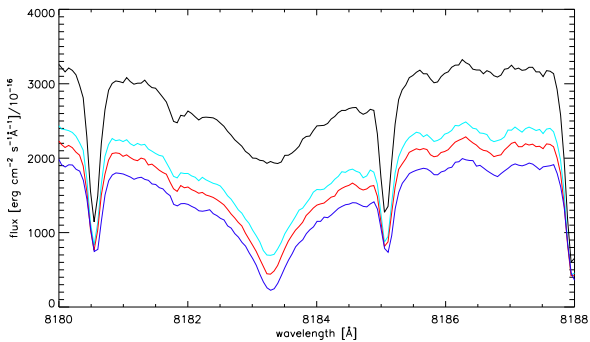}
\caption{\label{naline} \ion{Na}{i} line at 8183 \AA. The quiescent spectrum no. 1
is shown in blue/lowest line. Colour coding as in Fig~\ref{kline}.
In contrast to the \ion{K}{i} line in Fig. \ref{kline}, no emission core
is seen, but the line fills in with chromospheric emission, resulting in an
emission line in the flare-only spectra.
}
\end{center}
\end{figure}

In the blue spectrum no. 3 and the red spectrum no. 9, we detected 1143 chromospheric
emission lines altogether, with 989 out of them located in 
the blue arm and 154 in the red arm. Out of the 154 emission
lines in the red arm spectrum, 110 were analysed in the flare-only spectrum.
While the blue arm typically comprises pure emission lines,
emission cores in absorption lines and filled in absorption lines are also found in the red arm.
From our red spectra, we show the Paschen 7 line (pure emission) in the flare spectra (Fig. \ref{paschen}), 
the \ion{K}{i} line (emission core) at 7699 \AA\,(Fig. \ref{kline}), and the 
\ion{Na}{i} line (filled in absorption) at 8183\AA\,(Fig. \ref{naline}) as three representative examples of each type of enhanced flare emission.  

To determine the line parameters we used
an IDL routine that automatically fits
background, central wavelength, FWHM, and line flux as free parameters
using a Gaussian parameterisation for every manually indicated emission line
in a certain wavelength range. We typically chose 10~\AA\, as the wavelength
range for one fitting process, since the background variations on such an
short wavelength interval are acceptable. The applied program was kindly 
provided to us by
Dr. Jan-Uwe Ness. The line fit parameters including a (possible) identification
can be found in Table \ref{linetable}. 

\begin{table*}
\caption{\label{linetable}First 5 rows of the line catalogue. The whole
table is accessible electronically via http://cdsweb.u-strasbg.fr/cgi-bin/qcat?J/A+A/.}  
\begin{tabular}[htbp]{ccccclrcc}
\hline\hline
&central wavelength & FWHM & flux & catalogued wavelength & ion & multiplet & id flag & comment \\
&[\AA] & [\AA] &[erg\,s$^{-1}$\,cm$^{-2}$] & [\AA]   & & & & \\
\hline
&3057.45 &   0.057 & 4.43e-13 &   3057.446 &   FeI  &   28 &  0 & n \\
&3059.09 &   0.057 & 4.27e-13 &   3059.086 &   FeI  &    9 &  1 & n\\
&3062.26 &   0.045 & 2.08e-13 &   3062.234 &   FeII &  108 &  0 & n\\
&3064.62 &   0.042 & 9.17e-14 &   3064.623 &   NiI  &   26 &  0 & n\\
&3065.32 &   0.042 & 1.34e-13 &   3065.315 &   FeII &   97 &  0 & n\\

\end{tabular}
\end{table*}

In the line catalogue we provide the measured emission line fluxes. These
may be affected by rather large
errors mainly for two reasons. First, the description  with a Gaussian
may result in a poor fit if the lines have broad wings. The 18 affected lines belong to the ions 
\ion{H}{i}, \ion{O}{i}, \ion{Fe}{i}, and \ion{He}{ii} and 
are flagged accordingly.
Second, the background/quasi-continuum may be somewhat ill-defined. This is 
true for emission
lines in a broader emission line wing (stated in the remarks), 
but also for emission
cores of absorption lines and for varying background, with the latter found 
especially in the red arm due to molecules. 
Also, the H$\alpha$ line and  the \ion{Ca}{ii} triplet lines are 
saturated in the peak flare spectrum, which was used for the identification list.
Moreover, 5 Balmer/Paschen lines could not be fitted at all due to their large width  
(but they are clearly present). Therefore, individual lines may have flux errors
larger than a factor of 2, but normally the flux measurements should have
errors below 30 percent. 

We mainly used the Moore catalogue \citep{Moore} for identifying the lines, 
a few lines were identified using the NIST database \footnote{Available online
under\\ \mbox{http://physics.nist.gov/PhysRefData/ASD/index.html}}. For the
identifications from the Moore catalogue, the multiplet is also given in the
catalogue. The spectra were first shifted to the rest wavelength for the identification 
process. This enables us to find systematic lineshifts caused by the flare, which was not the case in 
general. The average displacement between measured
central wavelength and laboratory wavelength is 0.01 \AA\, in the blue arm and 
0.05\AA\, in the red arm. 

We were not able to identify 7 emission lines. Another
167 (15 \%) lines have only tentative identifications for one of the following reasons:
(1) the line was not found in the Moore catalogue, (2) the wavelength shift to the 
possible laboratory wavelength is large, (3) the line is severely blended  with 
other lines, and (4) (most often) the line is the
only one out of the multiplet. However, there are few lines, which are 
single identifications
out of a multiplet or are from a singlet, that are flagged as secure 
identifications. The catalogue includes a set of flags providing information on the
identification status of the line. If there are other possible identifications
for the line, these are normally given as remarks, even if the line is
identified securely. The latter may be the case if the line belongs to a
multiplet out of which a number of lines are already found, but there are
other (plausible) identifications with identical or very near laboratory 
wavelength. 

Since we excluded doubtful features from our line list,
the list cannot be claimed to be complete, especially for the weak lines with
fluxes below $5\cdot 10^{-14}$ erg\,s$^{-1}$\,cm$^{-2}$. Also, we partly used the flare-only spectrum in the red arm, as denoted by the identification flag for individual lines. To create this flare spectrum,
we subtracted the first spectrum of the May 19/20 night, which shows CN~Leo
in quiescence. 

The identified lines belong to 35 different ions in total, and the 
basic statistics of these flare lines can be found in Table \ref{identification}. 
For the appearance of \ion{V}{ii} lines in emission,
we are not aware of any description of
\ion{V}{ii} chromospheric emission lines for flare stars in the literature.

\setlength{\tabcolsep}{4.5pt}
\begin{table}
\caption{\label{identification}Chromospheric emission line identification}  
  
\begin{tabular}[htbp]{ll|ll}
\hline\hline
ion & no. lines$ ^1$ & ion & no. lines$ ^1$  \\

\hline

 \ion{H}{i}  & 15(15)&  \ion{Mg}{i} & 13(11)\\
 \ion{He}{i} &  7(5) &  \ion{Mg}{ii}&  5(3) \\
 \ion{He}{ii}&  1(1) &  \ion{Na}{i} &  2(1) \\
 \ion{Sc}{ii}& 16(16)&  \ion{Na}{ii}&  0    \\
 \ion{Ti}{i} & 15(6) &  \ion{K}{i}  &  1(0) \\
 \ion{Ti}{ii}&125(116)& \ion{Ca}{i} & 13(11)\\
 \ion{V}{ii} & 32(30)&  \ion{Ca}{ii}&  9(8) \\   
 \ion{Cr}{i} & 16(12)&  \ion{C}{i}  &  5(5) \\
 \ion{Cr}{ii}& 72(59)& 	\ion{N}{i}  &  1(0) \\	       
 \ion{Mn}{i} & 13(11)& 	\ion{O}{i}  &  6(6) \\      
 \ion{Mn}{ii}&  9(8) &   \ion{Ne}{i} &  1(0) \\ 
 \ion{Fe}{i} &494(428)&  \ion{S}{i}  &  2(0) \\
 \ion{Fe}{ii}& 90(70)&    \ion{Al}{i} &  4(2)\\ 
 \ion{Co}{i} & 47(43)& 	  \ion{Si}{i} & 10(3)\\ 	      
 \ion{Co}{ii}&  1(1) & 	  \ion{Si}{ii}&  1(0) \\	      
 \ion{Ni}{i} & 98(87)& 	  no id       &  7    \\      
 \ion{Ni}{ii}&  5(3) &   & \\
 \ion{Cu}{i} &  3(2) &   & \\
 \ion{Ba}{ii}&  1(1) &   & \\
 \ion{Y}{ii} &  1(0) &   & \\
 \ion{Zr}{ii}&  2(2)&    & \\

\end{tabular}\\
$^{1}$ in parenthesis the number of secure identifications\\
\end{table}

\subsection{Comparison to other work}

Many of the emission lines in the flare-only
spectrum (with the quiescent spectrum subtracted) have not been observed
before. Flare stars are traditionally observed in 
the ultraviolet and X-ray regime, rather than at these red wavelengths.  In addition, observations in this wavelength
range often lack the resolution necessary for resolving the metal emission
lines without wings. Nevertheless, most of the stronger emission
lines like the Paschen lines, the \ion{He}{i}, the \ion{Ca}{ii} lines, and others
have been described in spectacular flares on very late-type M dwarfs. For example,
2MASS J0149090+295613 (spectral type M9.5 V) was observed by \citet{2Massflare} 
during a flare showing 16 emission lines in the red part of the spectrum 
among them 5 Paschen lines and the \ion{Ca}{ii} triplet.  
\citet{Schmidt} also describes two flare events on LP 412-31 (M8) and 
2MASS J1028404-143843 (M7), showing emission lines in the considered wavelength range. 
Both stars show the \ion{Ca}{ii} triplet and hydrogen Paschen emission up to
P$_{11}$. During all these three flares, the Paschen
lines are much more prominent than in our case. We ascribe this to the
earlier spectral type of CN~Leo, which has more continuum flux at these
wavelengths. 

As another example we compare our observations to flares on the old inactive M4 star 
Gliese~699 (Barnard's star) by \citet{Barnard}. There, the wavelength coverage was 
from 3700 to 10\,800~\AA, with a spectral resolution of $\sim$ 60\,000.
\citet{Barnard} have also produced an emission line list. In the overlapping
wavelength range with our blue arm (from 3730 to 3860~\AA) they find 28
lines, where we find 135 emission lines. Among these, there is one line at 3772
\AA\ where we located \ion{Ni}{i} at 3772.530~\AA, while \citet{Barnard}
identified \ion{Fe}{i} at 3772.23~\AA. With the exception of this one line, 
we found all lines
also identified by \citet{Barnard} in the blue wavelength range. In the red
arm, we found 154 lines, while \citet{Barnard} detected 11 enhanced lines redward
of 6400~\AA. Again, we were able to locate all of their lines with one discrepancy at
6462~\AA, where we found a line at 6462.566~\AA\ tentatively identified as
\ion{Ca}{i}, while \citet{Barnard} found an enhanced line at 6462.73~\AA\ 
identified as an \ion{Fe}{i} line. It is not clear why we do not see these two
\ion{Fe}{i} lines identified for Barnard's star. Our resolution is sufficient
to detect these two lines and not to confuse them with our additional detections.

\subsubsection{\ion{He}{ii} emission from the transition region}

Among the chromospheric emission lines, there is an \ion{He}{ii} line at 3203.104~\AA\
originating from the transition region. This line, which is barely recognisable outside of flares, reacts to the flare as well, but not as strong and fast as the
chromospheric lines. During the flare peak in spectrum no.~3, the \ion{He}{ii}
line is not as strong as the two flanking chromospheric \ion{Ti}{ii}
emission lines. In the following spectrum no.~4, the two \ion{Ti}{ii}
lines have completely decayed, while the \ion{He}{ii}
line is still present. As found by \citet{fuhrmeister_liefke}, the 
situation is quite different
during the flare onset of 13 December 2005 as seen in spectrum
no.~12. There, the \ion{He}{ii} line reacts
much more strongly than the two \ion{Ti}{ii} lines,
which may be interpreted as due to different energy deposition heights for
the two flares. The strong flare discussed here had an energy deposition into
very deep layers (as can also be seen from the strong blackbody/photospheric
reaction and the enormous width of the Balmer series lines). Thus, the transition region seems to be heated more gently,
but of course this is difficult to decide from only one line. 

\section{Line asymmetries}\label{asym}

Line asymmetries during flares have been described at various times for
different emission lines. Blue excess-line emission was described
for example by \citet{Gunn} for a flare on AT~Mic, while
 \citet{Crespo} note red line asymmetries
in the Balmer lines during various flares on AD~Leo (M3V). The latter found the
asymmetries stronger for larger flares and stronger for higher members
of the Balmer series. They also analysed the \ion{Ca}{ii} H \& K lines, but
 did not find any asymmetry. Also, \citet{LHS2034} found red wing asymmetries in
the Balmer and \ion{He}{i} lines for a long duration flare on LHS~2034 (M6V).
The additional red wing flux decayed faster for the Balmer lines than
the line flux. For the \ion{He}{i} lines, the flux creating the asymmetry
first increased relative to the line flux and then decayed; examples of
papers describing line asymmetries are given by these authors.

Additional blue wing flux is normally interpreted as upmoving material, which
is expected for stellar flares, while additional red wing flux has been ascribed 
mostly to downflows during the
flares. For the Sun, similar mass flows (chromospheric downward condensations)
are known \citep{Canfield} to last for a few minutes. Since the asymmetries in
M dwarfs often persist much longer, various flare kernels producing several
downflows have been proposed \citep{Doyle,LHS2034}.

In the present data, line asymmetries have also been found. Besides
asymmetries in the Balmer and \ion{He}{i} lines, asymmetries for 
\ion{Ca}{ii} lines have been found for the first time. Asymmetries may 
be present in the \ion{Mg}{ii} lines, but they are all very weak,
so that a red wing is hard to distinguish from the noise/molecular lines. 

In the following we give a detailed presentation of the line asymmetries.

\subsection{\ion{He}{i} lines}

\begin{figure*}
\begin{center}
\includegraphics[width=8cm]{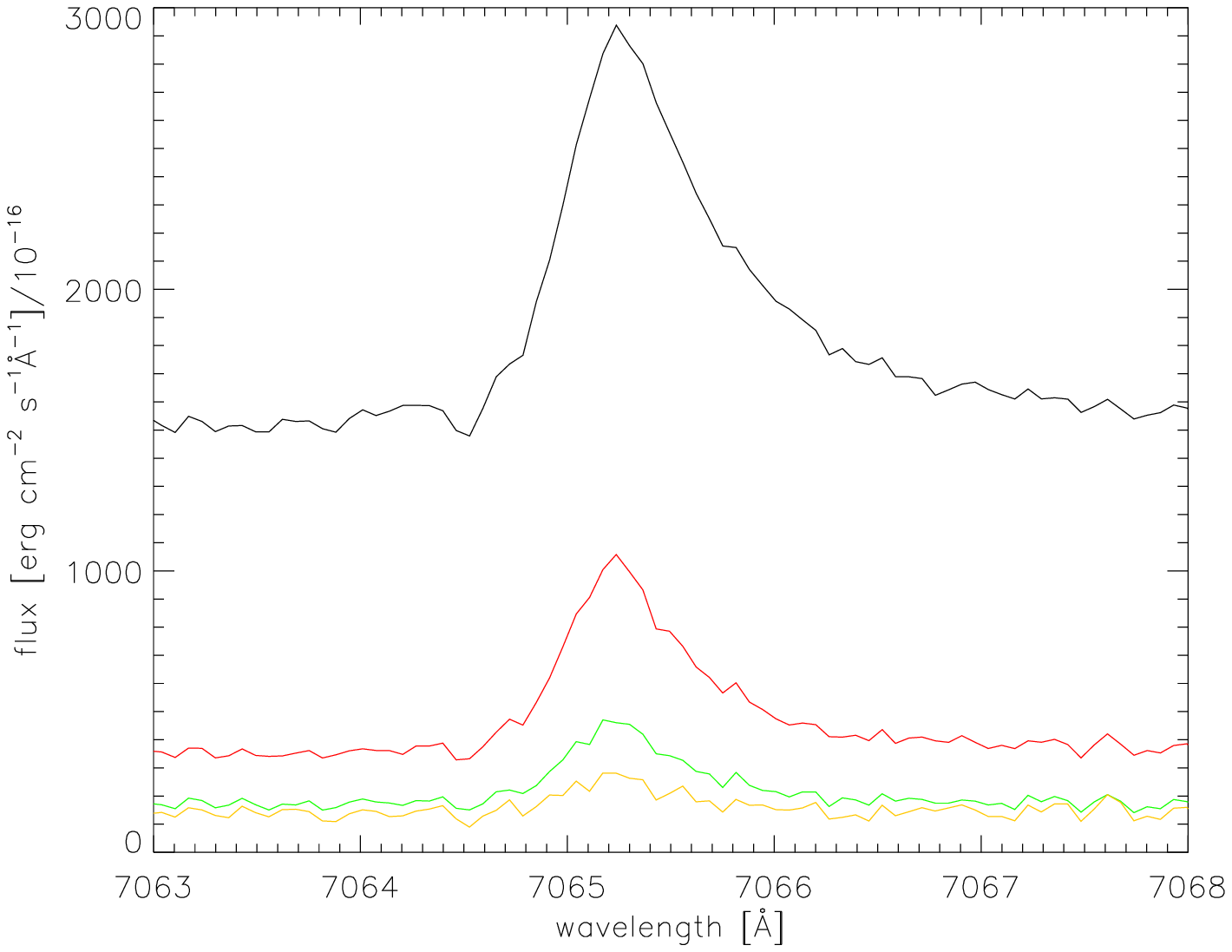}
\includegraphics[width=8cm,height=6cm,clip=]{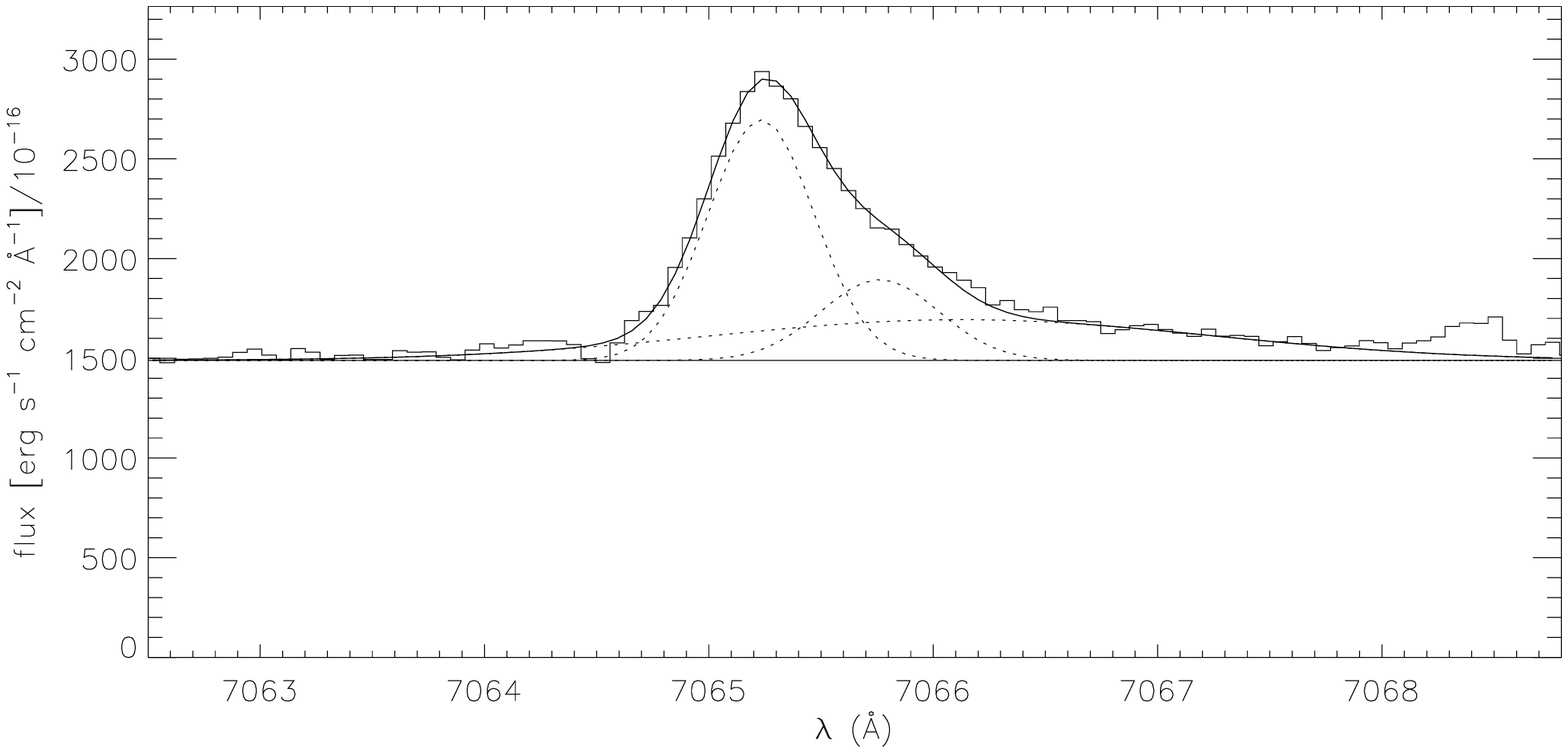}
\caption{\label{Heasym} Red-wing asymmetry for the \ion{He}{i} line at
7065.30/7065.93~\AA. Left: Flare spectrum no.~9 of the line and the three 
consecutive spectra (with the quiescent spectrum subtracted for each
spectrum). Right: Three-component Gaussian fit of the line with CORA.
The asymmetry manifests itself in the broad Gaussian component. The
background has been fitted manually.
}
\end{center}
\end{figure*}

Seven \ion{He}{i} lines are observed in the flare peak spectra no.~9 (red arm)  and 3 (blue arm).
Five lines clearly show additional flux in the red wing, one is located in a 
wavelength region too crowded to recognise any asymmetry, and one line 
is very weak, but looks slightly asymmetric.  
An example of these asymmetries can be found in Fig.~\ref{Heasym}
for the double \ion{He}{i} line at 7065~\AA. 

We also fitted the two strongest flare-only
(quiescent spectrum removed) \ion{He}{i} lines
with CORA \citep{cora}. An example of the fit quality can also be found in
Fig.~\ref{Heasym}. We fitted the double line at 7065 \AA\, with two Gaussian
components plus one  broad Gaussian for the additional line flux in 
the asymmetry. For each Gaussian the free fit parameters are central wavelength,
FWHM, and amplitude. The background was fitted manually, since the line is very 
broad and therefore 
contributes significantly to the background as fitted by CORA. The same fit procedure was applied
to the line at 6678~\AA, but with the background fitted automatically and only
one main component. The fit results can be found in Table~\ref{HeI3}.

\begin{table*}
\caption{\label{HeI3}Line-fit parameters of the \ion{He}{i}  
 line at 7065 and 6678 \AA.}  
\begin{tabular}[htbp]{cccccc}
\hline\hline
 \multicolumn{6}{c}{\ion{He}{i} at 7065 \AA$^1$}\\
No. & flux narrow 1&flux narrow 2& flux broad & FWHM broad & velocity\\
  &\scriptsize{[$10^{-16}$ erg\,s$^{-1}$\,cm$^{-2}$]}  &\scriptsize{[$10^{-16}$ erg\,s$^{-1}$\,cm$^{-2}$]} & \scriptsize{[$10^{-16}$ erg\,s$^{-1}$\,cm$^{-2}$]} &[\AA] &  km\,s$^{-1}$ \\
\hline

8 & 1294 $\pm$118 & -            & 1212 $\pm$192& 0.85 $\pm$0.09& -3.8$\pm$11.9\\
9 & 11143 $\pm$206& 4139 $\pm$223& 8844 $\pm$442& 1.11 $\pm$0.03& 38.6$\pm$19.5\\
10 & 5562 $\pm$112 & 1386 $\pm$101& 2112 $\pm$192& 1.04 $\pm$0.10&45.8$\pm$13.1\\
11 & 2389 $\pm$78  & 437  $\pm$62 & 908 $\pm$154 & 1.13 $\pm$0.01&28.0$\pm$19.5\\

 \multicolumn{6}{c}{\ion{He}{i} at 6678 \AA$^2$}\\
\hline
8& 1646 $\pm$ 145 & &2235 $\pm$ 217& 0.66 $\pm$0.03&-10.8$\pm$11.2\\
9& 22360$\pm$ 295 & &24583 $\pm$467& 1.02 $\pm$0.01& 24.7$\pm$0.5 \\
10& 10033$\pm$ 150& &6512 $\pm$207 & 0.91 $\pm$0.05&22.4$\pm$2.7  \\
11& 3993$\pm$ 105 & &2310 $\pm$140 & 0.68 $\pm$0.07&10.8$\pm$14.8 \\
12& 2022$\pm$ 75  & &610 $\pm$110  & 0.84 $\pm$0.08&9.0$\pm$25.1  \\
13& 628$\pm$ 46   & &              &               &              
\end{tabular}\\
$ ^1$ The first narrow component of the line is stable at 7065.19~\AA,
the second is stable at 7065.73~\AA. Only for spectrum no. 8, the line center is
shifted to 7065.10$\pm$ 0.12~\AA. The FWHM of both narrow components
is $\sim$0.23~\AA. The background is fitted manually.  
$ ^2$ The narrow component of the line is stable at 6678.15~\AA\ besides for spectrum no. 8, where it is
blue-shifted to 6678.03$\pm$0.10~\AA.
\end{table*}

In the spectrum no.~8, covering the flare onset, the two strongest \ion{He}{i} 
lines at 6678 and 7065~\AA\ start to react to the flare. Both lines are
blue-shifted and exhibit additional broad, blue-shifted components.   
For the line at 7065~\AA, we did not find the second narrow component
in the data, although this may be due to the low flux in the line. 
No blue asymmetries can be found in the blue arm of the spectrum, since the
blue spectrum no 2 covering the flare onset is dominated by the continuum and the 
\ion{He}{i} lines are too weak against this strong background.

\subsection{\ion{Ca}{ii} lines}

During the flare, we found 9 \ion{Ca}{ii} lines, out of which 5 are located 
in the blue arm. Seven lines show additional red wing flux, one 
is clearly blended with an \ion{Fe}{i} line, and another one is so weak that no
asymmetry -- if present -- can be recognised. 
The \ion{Ca}{ii} triplet is saturated in the
line cores in spectrum no. 9, but the wings are not affected. All three lines show red
asymmetries in the wings, although not as strong as the \ion{He}{i} lines.
We fitted the three lines in the flare-only spectra, again using a narrow and a broad Gaussian, bearing 
in mind that the narrow component flux is saturated. Nevertheless, the
line shifts are unaffected by the saturation. Our line fit parameters are
given in Table~\ref{catriplett}. After spectrum no.~11, the broad component vanishes.
In spectrum no.~8, covering the flare onset, no broad component is present.
However, all three lines are blue-shifted, although only within 1 to 2 
$\sigma$ (see Table \ref{catriplett}).

\begin{table*}
\caption{\label{catriplett}Line shift of the broad component of the \ion{Ca}{ii} triplet lines from the flare-only spectra.}  
\begin{tabular}[htbp]{cccccccccc}
\hline\hline
     &\multicolumn{3}{c}{\ion{Ca}{ii} at 8498.018 \AA$ ^1$}&\multicolumn{3}{c}{\ion{Ca}{ii} at 8542.089 \AA$ ^2$}&\multicolumn{3}{c}{\ion{Ca}{ii} at 8662.140 \AA$ ^3$}\\
No.  & position & velocity & FWHM & position & velocity & FWHM & position & velocity & FWHM \\
   & [\AA] & [km\,s$^{-1}$] & [\AA] & [\AA] & [km\,s$^{-1}$] & [\AA]&[\AA]& [km\,s$^{-1}$] & [\AA]\\
\hline

9 &8498.32$\pm$0.02& 11.6$\pm$0.7& 0.89$\pm0.04$&8542.22$\pm$0.13& 5.3$\pm$ 4.6& 1.00$\pm$0.01&8662.33$\pm$0.01&7.3$\pm$0.3 &0.91$\pm$0.02 \\ 
10 &8498.48$\pm$0.10 & 17.3$\pm$3.5&0.72$\pm$0.02&8542.17$\pm$0.17&3.5$\pm$ 6.0&0.78$\pm$0.02&8662.23$\pm$0.19&3.8$\pm$6.6 &0.70 \\ 
11 &8498.34$\pm$0.43 & 12.3$\pm$15.2&0.72$\pm$0.02&8542.07$\pm$0.15&0.0$\pm$5.3 &0.78$\pm$0.05&8662.22$\pm$0.22&3.4$\pm$7.6 &0.70 \\

\end{tabular}\\
$ ^1$ The narrow component is stable at 8497.99$\pm$0.02 \AA, besides spectrum no.8 which is blue-shifted to 8497.80$\pm$0.11 \AA\, (-6.7 $\pm$ 3.9 km\,s$^{-1}$).\\
$ ^2$ The narrow component is stable at 8542.07$\pm$0.03 \AA, besides spectrum no. 8 which is blue-shifted to 8541.96$\pm$0.11 \AA\, (-3.8 $\pm$ 3.8 km\,s$^{-1}$).\\
$ ^3$ The narrow component is stable at 8662.12$\pm$0.04 \AA, besides spectrum no. 8 which is blue-shifted to 8662.02$\pm$0.07 \AA\, (-3.8 $\pm$ 2.4 km\,s$^{-1}$).\\
\end{table*}

\subsection{\ion{H}{i} lines}

Asymmetries have also been found for the H$\alpha$ line. 
During the flare onset in spectrum no.~9, the
line has a footpoint width of about 70~\AA\ and the inner wings
are also saturated; therefore,  we refrained from using this spectrum for the analysis. 
In the decay phase, the core is saturated for spectra no.~10 to 14, but the wings, where the 
asymmetries are
found, are not affected. We fitted the net-flare line (with the quiescent spectrum
no.~1 subtracted) with CORA using two Gaussians, which
gives a good empirical description of the line. 

\begin{figure}
\begin{center}
\includegraphics[width=8cm]{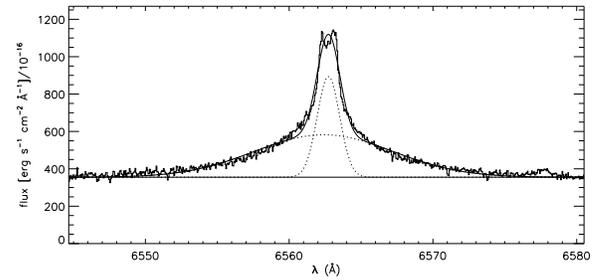}
\caption{\label{halpha} The H$\alpha$ line from spectrum no. 8 with the quiescent
spectrum subtracted. The line exhibits a clear blue asymmetry due to the
blue-shifted broad component. The narrow component is also slightly blue-shifted. 
}
\end{center}
\end{figure}

The H$\alpha$ line starts to react to the flare in spectrum no. 8, where
it develops broad wings. Moreover, the narrow component is slightly
blue-shifted  at 6562.71$\pm$0.06~\AA,
whereas the broad component is clearly blue-shifted 
leading to a blue asymmetry, as can be seen in Fig.~\ref{halpha}. 
The line core (note, that spectrum no 8 is not 
saturated) exhibits a self-reversal due to non-LTE effects, 
which is typical for dMe stars and
also reproduced in chromospheric modelling of M dwarfs; see e.\,g.
 \citet{Cram, Robinson}, and \citet{Houdebine1}. The weak asymmetry of the red and the blue
peaks also points to the complicated dynamical processes during flares
and may be due to a wave propagating downwards. Examples of theoretical
H$\alpha$ profiles from radiative hydrodynamic flare models for M dwarfs
also exhibiting peak asymmetries due to propagating waves
can be found in \citet{Allred}. 

The broad component is red-shifted in the
spectra covering the flare decay phase.
 The derived wavelength shifts are listed in Table \ref{Halphat},
assuming a rest wavelength of 6562.817~\AA\ \citep{Moore}. The
broad component has vanished in spectrum no. 15.  

\begin{table}
\caption{\label{Halphat}Line parameters for the broad component of the H$\alpha$ line. Note, that there
is no measurement for spectrum no. 9.}  
\begin{tabular}[htbp]{cccc}
\hline\hline
No.  & position &  velocity & FWHM \\
  & [\AA] & [km\,s$^{-1}$] & [\AA] \\
\hline

8 & 6562.47$\pm$ 0.10 & -15.8$\pm$ 4.6&4.82$\pm$0.02\\
10 & 6563.39$\pm$ 0.01& 26.2$\pm$0.5& 4.10 $\pm$0.01\\
11 & 6563.08$\pm$ 0.01& 12.0$\pm$0.5& 2.90 $\pm$ 0.01\\
12 & 6563.21$\pm$ 0.01& 17.9$\pm$0.5& 2.12 $\pm$ 0.02\\
13 & 6563.60$\pm$ 0.06& 35.8$\pm$2.7& 1.50 $\pm$ 0.09\\
14 & 6562.88$\pm$ 0.38& 2.9 $\pm$17.4& 3.10 $\pm$ 0.11\\

\end{tabular}
\end{table}   

Besides the H$\alpha$ line, we also analysed the H$_{9}$ line in detail. For the
even higher Balmer lines, we refrained from a detailed analysis, since the
broad wings of these lines have to be deblended from metal lines, Due to decreasing flux this becomes more and more difficult for higher Balmer 
lines. 

We performed a simultaneous fit with three metal lines for the H$_{9}$ line, i.\,e. 
a 5-component Gaussian fit with three Gaussians for the metal
lines and one narrow and one broad Gaussian for the H$_{9}$ line itself.
This procedure led to a very good description of the data, except for spectra no.~2
and 3. For these two spectra, we changed the narrow Gaussian profile
to a Lorentzian profile again leading to acceptable fits. For spectrum
no.~2, which covers the flare onset, a weak blue asymmetry could be detected with
the broad component at 3835.34$\pm$0.03~\AA. The Lorentzian describing the
narrow component is only blue-shifted within the error bars: 
3835.36$\pm$0.08~\AA. The FWHM is 0.51~\AA\ for the Lorentzian and 2.20~\AA\
for the broad Gaussian. During the
flare peak spectrum no.~3, the broad component is red-shifted at 3835.47$\pm$0.01~\AA\
 with a FWHM of 3.34~\AA.
The following spectra tend to have red-shifted broad components, but within the
large errors. Nevertheless, no line shift in the consecutive spectra agrees 
with what we found for H$\alpha$, where the broad component vanishes after spectrum
no. 14 (cf., Fig.~\ref{lightcurve}).  

In addition to the Balmer lines, we searched the Paschen lines for
asymmetries, but none were found; however, since these lines are
very broad and noisy, an asymmetry could easily be hidden in the data (see Fig.~\ref{paschen}). 

\subsection{Discussion of the line asymmetries}

We measured line asymmetries for \ion{He}{ii}, \ion{Ca}{ii}, and Balmer
lines during the flare, but not for the Paschen lines and other metal lines (although the \ion{Mg}{ii}
lines also may exhibit small asymmetries). While in the case of the Paschen lines this can
be attributed to the enormous line widths hiding asymmetries, this argument is not valid
for the metal lines. However, the lines of the above-mentioned ions tend to form 
at different chromospheric heights (and therefore at different temperatures)
than the rest of the observed metal lines. While
the line cores of the above-mentioned ions tend to form in the upper chromosphere 
or transition region onset, the line cores of the other metals tend 
to form in the lower to mid chromosphere 
\citep{Vernazza, Fuhrmeister}. This gives a hint at least to the temperature of the
moving material, which we ascribe the asymmetries to. The material must be rather hot
(about 8000 to 10\,000 K) compared to the rest of the chromosphere. 
 We also ascribe the non-uniform reactions
of the asymmetries of the different lines to
different line formation conditions.  Nevertheless, trends in
the reaction of the lines can be noticed:
In our spectra, we find asymmetries in the blue as well as in the red wings. Asymmetries
on the blue side are found during flare onset, red wing asymmetries during decay phase. This makes the interpretation as moving
material very straightforward. At the flare onset, the chromospheric evaporation
is seen as a blue shift (although partly only within the errors) of the main/narrow
components of the analysed strong lines. In addition, there seems to be a rising
`cloud' of material,
but moving with a different velocity producing the broad component. 
The main/narrow line component does not indicate any moving material 
in the decay phase, but the `cloud' is now
moving downward, leading to the observed red asymmetries.
For the H$\alpha$ line, we detected
not only a decay of the asymmetry, but also variations in the asymmetry from
spectrum to spectrum, favouring a scenario of not one single uniform downflowing 
'cloud', but of a multitude of downflows. These are de-accelerated when hitting
deeper and denser atmospheric layers, which is seen in the generally 
decreasing velocity
of the broad component with evolution of the flare.

During the decay phase, downflow
velocities range from $\sim$ 5 to 40 km\,s$^{-1}$, which is probably
supersonic. The sound velocity in models of quiescent chromospheres is
 about 10 km\,s$^{-1}$ \citep{Jevremovic}. Similar downflow
velocities have been measured for the flare on LHS 2034 \citep{LHS2034},
but downflow velocities of up to 60 km\,s$^{-1}$ are
normally observed for the Sun \citep{Fisher}. 
 
Interpretating the
FWHM of the broad component is difficult because it may have
different components. At least for the hydrogen lines, Stark broadening
should contribute to the line width, and turbulent motions are also a possible
interpretation. Since there may be different downflow kernels, the width
of the line may be due to the different bulk motions of these kernels that
are integrated in our spectra.

In conclusion, we tentatively interpret our observations in a scenario of a chromospheric 
prominence that is lifted during the flare onset and then raining down
during the decay phase.

\section{Discussion and conclusion}\label{concl}

We presented  the optical part of a multiwavelength campaign that
caught CN~Leo in a spectacular flare. There are only few observations
of such giant flares, but even these few examples show very different timescales
and lightcurve characteristics. While the CN~Leo flare lasted about
50 minutes, a 5-magnitude flare in the U-Band on UV~Cet lasted
only 12 minutes \citep{Eason}, while the large flare
on AD~Leo from 1985 lasted more than 4 hours \citep{Hawley_Pettersen}.
Compared to somewhat smaller (but nevertheless very large) flares, these
can be of the same duration or even longer than the CN~Leo flare. For example, 
a 2.1 magnitude flare in the U-band was observed on AD~Leo to last about 50
minutes \citep{AD-Leo1984}, while a large flare on AT Mic lasted  about 1.5
hours \citep{AT-Mic}. 

Most of the large flares show a double (or triple) peaked lightcurve in the U-band and
also for Balmer emission lines like the CN~Leo flare \citep{Hawley_Pettersen,
AD-Leo1984, AT-Mic}. The UV~Cet flare had an even more complicated substructure
with at least 4 peaks in the major outburst \citep{Eason}. This multiple-peak
flare pattern is usually interpreted in terms of different reconnection events,
which may trigger each other in succession. The CN~Leo flare shows two major
peaks that can be easily noticed in the exposuremeter data  in Fig.~\ref{lightcurve}, with a minor peak between the
two. The time between the two major peaks is about 200 seconds.
For the 2.1 mag flare on AD~Leo,
the two major peaks are separated by about 9 minutes \citep{AD-Leo1984}.
For all other mentioned flares (besides UV~Cet), the separation is even longer.

The flux peaks of chromospheric lines are typically delayed with respect to the U-band flux peak,
with the \ion{Ca}{ii} H and K lines even more delayed than the Balmer lines
\citep[e.\,g.][]{AT-Mic, Hawley_Pettersen}. The time resolution of our spectra
is unfortunately too low to observe any delay in the peak, but as is apparent
in Fig.~\ref{line_flux}, the flux of the \ion{Ca}{ii} triplet line
decays more slowly than the other lines. This behaviour is compatible with what is
normally observed for the \ion{Ca}{ii} H and K lines, which decay more gradually
than Balmer and Helium lines \citep[e.\,g.][]{AD-Leo1984}, consistent with an
interpretation of a faster evolution of high-temperature lines 
in the chromosphere \citep{Houdebine3}.

The optical spectral data allowed us to trace
the chromospheric reaction to the flare (with one \ion{He}{ii} transition
line covered). We found a wealth of chromospheric emission lines that has
not been observed before, especially in the red part of the spectrum; also, \ion{V}{ii} and \ion{Sc}{ii} lines show up in
the blue arm. Especially in the flare onset spectrum no.~2,
the higher Balmer lines develop very broad damping wings that can only  be
fitted involving a Lorentzian profile. With respect to line shapes and profiles,
we  find asymmetries varying from blue excess flux during flare onset
to red excess flux during the decay phase. In the red excess for H$\alpha$, not only 
a general 
decay is found, but also variations from spectrum to spectrum, providing strong
evidence for a scenario involving multiple flare kernels, each triggering 
chromospheric downward condensations. We ascribe the blue excess flux during flare
 
The optical quasi-continuum is also quite massively affected by the flare.
The blue spectrum is clearly blackbody-dominated during flare onset. The
same is true for the red spectrum with the quiescent spectrum removed.
The derived temperatures for the gas producing the blackbody radiation
in the first few seconds of the flare onset exceed of 20\,000 K and 
then rapidly cool. 

This set of excellent optical data, which characterises the photosphere and
chromosphere of CN~Leo during a giant flare is suited well to classical
semi-empirical chromospheric flare modelling with a stellar atmosphere code.
Modelling using a multitude of emission lines will be presented in
a forthcoming paper in this series \citep{paper3}.
\begin{acknowledgements}
  B.F. and C.L. acknowledge financial support by the DLR under
  50OR0105. A.R. has received research funding from the DFG as an Emmy
  Noether Fellow (DFG 1664/4-1).
\end{acknowledgements}

\bibliographystyle{aa}
\bibliography{AA9379b}

\end{document}